\documentclass[aps,a4paper,preprint,12pt,preprint,prb]{revtex4}

\usepackage{graphicx}
\usepackage{amssymb}
\usepackage{epstopdf}
\usepackage{natbib}

\begin{document}
\title{Gas-liquid phase separation in oppositely charged colloids: stability and interfacial tension}
\author{Andrea Fortini}
\author{Antti-Pekka Hynninen\footnote{Present address:
Department of Chemical Engineering
Princeton University, Princeton NJ, 08544, USA}}
\author{Marjolein Dijkstra}
\affiliation{Soft Condensed Matter, Utrecht University,
   Princetonplein 5, 3584 CC Utrecht, The Netherlands.}

\begin{abstract}
We study the phase behavior and the interfacial tension of the
screened Coulomb (Yukawa) restricted primitive model (YRPM) of
oppositely charged hard spheres with diameter $\sigma$ using Monte
Carlo simulations. We determine the gas-liquid and gas-solid phase
transition using free energy calculations and grand-canonical
Monte Carlo simulations for varying inverse Debye screening
length $\kappa$. We find that the gas-liquid phase separation is
stable for $\kappa \sigma \leq 4$, and that the critical
temperature decreases upon increasing the screening of the
interaction (decreasing the range of the interaction). In
addition, we determine the gas-liquid interfacial tension using
grand-canonical Monte Carlo simulations. The interfacial tension
decreases upon increasing the range of the interaction. In
particular, we find that simple scaling can be used to relate the
interfacial tension of the YRPM to that of the restricted
primitive model, where particles interact with bare Coulomb
interactions.
\end{abstract}
\maketitle

\section{Introduction}
Coulombic interactions are important in a wide variety of physical
systems such as electrolytes, molten salts, plasmas, colloidal
suspensions, micelles, microemulsions, and  liquid metals. The
screened Coulomb (Yukawa) potential arises naturally for charged
particles in the presence of a screening distribution of
microions. The phase behavior of a pure system of hard spheres
interacting with screened Coulomb potentials has been well-studied
and the phase diagram displays stable fluid, face-centered-cubic
(fcc), and body-centered-cubic (bcc) crystal phases.\cite{Robbins1988,Meijer1997,ElAzhar2000,Hynninen2003} 
In this paper, we study  a binary fluid of oppositely charged particles
using computer simulations. While the phase diagram of the
restricted primitive model (RPM), consisting of a binary mixture
of equally sized  hard spheres carrying opposite charges of equal
magnitude, and interacting with bare Coulombic interactions, has
been widely studied,\cite{Orkoulas1994, Vega1996, Smit1996,
Bresme2000a, Abascal2003, Vega2003, Kalyuzhnyi2005, Fisher2005}
there is little information available on the phase diagram of the
Yukawa restricted primitive model (YRPM), where the hard spheres
of diameter $\sigma$ interact with screened Coulomb potentials
$u_{ij}=\pm \epsilon \sigma \exp[ -\kappa (r_{ij}-\sigma) ] /
r_{ij}$, with $r_{ij}$ the distance between particles $i$ and $j$,
$\epsilon$ the contact value, and $\kappa$ the Debye screening
parameter  (inverse of the Debye screening length). Recently,
\citet{Hynninen2006} determined the full phase diagram of the YRPM
for a  screening parameter $\kappa \sigma= 6$. At high
temperatures, the system behaves like a pure hard-sphere system,
with a transition between a fluid and a substitutionally
disordered fcc phase, where the opposite charges are distributed
randomly on a fcc lattice. At lower temperatures, a dilute gas
phase coexists with a high density CsCl solid phase, and the
gas-liquid transition is metastable with respect to freezing. At
high densities, various solid-solid transitions appear, e.g., a
transition from  CsCl to CuAu  and  from CuAu to the tetragonal
phase. Overall the system exhibits a phase behavior in striking
similarity with the RPM phase diagram,\cite{Smit1996, Bresme2000a,
Abascal2003, Vega2003, Hynninen2006} which displays a
fluid-disordered fcc transition at high temperatures, a stable
gas-liquid transition at low temperatures, and a fluid-solid
transition at higher densities. Since the RPM is the limit of the
YRPM for $\kappa\sigma \rightarrow 0$, we expect a crossover from
a metastable to a stable gas-liquid transition for
$0<\kappa\sigma<6$.

The gas-liquid transition for a similar model with pair potential
$u_{ij}=\epsilon \exp[ -\kappa (r_{ij}-\sigma) ]$, has been
studied using computer simulations.\cite{Caballero2004,Caballero2005,Caballero2005a} 
This interaction potential differs by a factor $\sigma/r_{ij}$ from our
model. The factor $\sigma/r_{ij}$ is of the order of unity, and we
expect the effect on the phase behavior to be small. This allows
us to  compare the results of Refs.
\cite{Caballero2004,Caballero2005,Caballero2005a} with the results
of the YRPM. In particular, \citet{Caballero2005} investigated the
critical temperature as a function of the screening parameter. 
In a later paper,\cite{Caballero2005a} the stability
of the gas-liquid separation with respect to the gas-solid
transition was estimated by computing the melting density of the CsCl structure.
This technique overestimate the stability of the gas-liquid binodal, with respect to our free energy calculations.
We will discuss the relationship between our results and those of Ref. \cite{Caballero2005a}  in more detail in Sec. IV.

An experimental realization of the YRPM is provided by
charge-stabilized  colloidal suspensions. Recently, it was shown
experimentally that the charge on the colloids can be tuned in
such a way that oppositely charged colloids can form large
equilibrium ionic colloidal crystals.\cite{Leunissen2005,
Bartlett2005, Hynninen2006, Hynninen2006b} 
Experiments, theory, and simulations based on screened Coulomb interactions are in good
agreement. The system studied in Refs.\cite{Leunissen2005,
Hynninen2006} had a Debye screening parameter $\kappa \sigma \sim
7$, and a gas-liquid phase separation was not observed.

The critical temperature and structure of the YRPM has also been
studied using integral equation theory,\cite{Carvalho1997} but no
information on the stability of the gas-liquid transition with
respect to freezing has been given.

The purpose of this paper is to determine the value of the
screening parameter $\kappa\sigma$ at which the gas-liquid
transition becomes stable for oppositely charged colloids. To this
end, we perform Monte Carlo simulations to compute the Helmholtz
free energies of the fluid and solid phases of the YRPM. We also
study the dependence of the critical parameters and the gas-liquid
interfacial tension on the interaction range. The critical
parameters and the values of the interfacial tension are
calculated using histogram reweighting methods and grand-canonical
Monte Carlo simulations. The paper is organized as follows. In
Sec. II, we describe the model, and we present the simulation
methods in Sec. III. The results are discussed in Sec. IV and we
end with some concluding remarks in Sec. V.

\section{Model}
We investigate the Yukawa restricted primitive model (YRPM)
consisting of $N$ spherical particles with a hard-core diameter
$\sigma$ in a volume $V$. Half of the spheres carry a positive
charge and the other half a negative charge of the same magnitude.
The pair interaction reads
\begin{equation}
\beta u(r_{ij})=\left \{
\begin{array}{ll}
\infty & r_{ij}\leq \sigma \\
 \pm \displaystyle \frac{\large \epsilon_{\rm Y}}{\large k_BT}
  \frac{\exp[ -\kappa (r_{ij}-\sigma) ]}{r_{ij}/\sigma} &  \sigma<r_{ij} < r_{\text{cut}}   \\
 0 & \text{otherwise}
\end{array}
\right. ,
\label{Eq:yu}
\end{equation}
where $r_{ij}$ is the distance between spheres $i$ and $j$,
$\kappa$ the screening parameter, $\beta \equiv 1/k_BT$ the
inverse temperature with $k_B$ the Boltzmann constant and $T$ the
temperature, and $\epsilon_{\rm Y}$ the contact value of the
potential. The cut-off value is $r_{\text{cut}}=3.6\sigma$. The
interaction is attractive for oppositely charged spheres, and
repulsive for like-charged spheres. We define a reduced
temperature $T_{\rm Y}^*=k_B T /\epsilon_{\rm Y}$ and measure
particle density in terms of the packing fraction $\eta=(\pi
\sigma^3 /6 ) N/V$.

According to the Derjaguin-Landau-Verwey-Overbeek (DLVO) theory,
\cite{Derjaguin1941, Verwey1948} the effective pair potential
between two charged spheres carrying the same number $Z$ of
elementary charges $e$ suspended in a sea of salt ions with
density $\rho_s$ is given by Eq. (\ref{Eq:yu}) with a contact
value
\begin{equation}
\frac{\epsilon_{\rm Y}}{k_BT} =\frac { Z^2 } {  \left ( 1+ \kappa
\sigma/2 \right )^2}\frac{\lambda_B}{ \sigma}\ . \label{dlvo}
\end{equation}
The Debye screening parameter reads $\kappa=\sqrt{8 \pi \lambda_B
\rho_s}$, where $\lambda_B= e^2/\epsilon_s k_B T$ is the Bjerrum
length and $\epsilon_s$ is the dielectric constant of the solvent.
It must be noted that, more recent theories on same charged colloidal spheres suspended in a sea of salt ions yield potentials of the form of
Eq.~(\ref{Eq:yu}), but with screening parameters that depend on
the charged colloid concentration.\cite{Graf1998, Dijkstra1998,
Roij1999, Denton2000, Warren2000, Belloni2000, Zoetekouw2005}
However, the exact functional form is yet unknown
\cite{Hynninen2005b} and different theories predict varying
functional forms. 
Furthermore, the DLVO theory was not originally derived for
oppositely charged spheres, but it can be extended using the linear
superposition approximation (LSA) to obtain the potential given by
Eqs.~(\ref{Eq:yu}) and (\ref{dlvo}).\cite{Bell1970} 
The extended DLVO theory has been shown to give good agreement with Poisson-Boltzmann \cite{MaskalyThesis} and primitive model
calculations \cite{HynninenThesis} at small $\kappa\sigma$, justifying the use potential in Eq.~(\ref{Eq:yu}) with the contact
value given by Eq.~(\ref{dlvo}). 
We will refer to the DLVO theory extended by the LSA simply as the
DLVO theory. To facilitate the comparison between the results of
the DLVO theory for different screening lengths $\kappa$, we
define a reduced temperature $T_{\rm C}^*=\sigma /Z^2 \lambda_B$
that is independent of $\kappa$ and equal to the the definition of the reduced temperature of the RPM.

\section{Simulation Methods}
In order to determine the stable phase for a given state point, we
compute the Helmholtz free energy as a function of $\eta$ and
$T^*_Y$. As the free energy cannot be measured directly in a
Monte Carlo simulation, we use thermodynamic integration
\cite{Frenkel2002,Frenkel1984,Polson2000} to relate the free
energy of the YRPM system to that of a reference system, whose
free energy is known. In the thermodynamic integration of the
fluid phase, we use the hard-sphere fluid as the reference state,
whereas in the solid phase, the reference state is the Einstein
crystal. We use a 10-point Gaussian quadrature for the numerical
integrations and the ensemble averages are calculated from runs
with 40000 MC cycles (attempts to displace each particle once),
after first equilibrating the system during 20000 MC cycles.
Employing a common tangent construction on the fluid free energy
density curves as a function of $\eta$, we find the points of
tangency that correspond to the densities of the coexisting gas
and liquid phase. A similar common tangent construction is used to
determine the coexistence between the fluid and solid phases, and
to check whether the gas-liquid separation is stable with respect
to the fluid-solid phase coexistence. In addition, we perform a
more detailed study of the gas-liquid binodal using methods based
on histogram reweighting. To this end, we employ grand-canonical
Monte Carlo simulations with successive umbrella sampling
\cite{Virnau2004} to overcome the free energy barrier between the
liquid and gas phase. In the successive umbrella sampling method,
the probability $P(N)|_{z_+,z_-}$ of having $N$ particles at
fugacity $z=z_+=z_-$ in a volume $V=L^3$ is obtained by sampling
successively 'windows' of particle numbers. In each window, the
number of spheres $N$ is allowed to fluctuate by one particle,
i.e., between 0 and 1 in the first window, 1 and 2 in the second
window, etc. We choose at random whether to make an attempt to
insert or to remove a particle such that, on average, the system
is charge neutral. The sampling of the probability ratio
$P(N)/P(N+1)$ is done, in each window, until the difference
between two successive samplings of the probability ratio is
smaller than $10^{-3}$. At phase coexistence, the (normalized)
distribution function $P(N)|_{z}$ becomes bimodal, with two
separate peaks of equal area for the liquid and gas phase. To
determine phase coexistence we calculate the average number of
particles
\begin{equation}
\langle N \rangle= \int_{0}^{\infty}N P(N)|_{z} dN.
\end{equation}
Subsequently, we use the histogram reweighting technique
\cite{Ferrenber1988} to determine the fugacity $z'$ for which the
equal area rule
\begin{equation}
\int_{0}^{\langle N \rangle }P(N)|_{z'}dN = \int_{\langle N
\rangle }^{\infty} P(N)|_{z'} dN,
\end{equation}
which is the condition for phase coexistence, is satisfied.

The gas-liquid interfacial tension $\gamma_{\lg}$ for a finite
system with volume $V = L^3$ is obtained from $P(N)|_{z'}$ at
coexistence:
\begin{equation}
\beta \gamma_{\lg,L} =\frac{1}{2 L^2}\left\lbrack \ln
\left(\frac{P(N_{\text{max}}^g )+P(N_{\text{max}}^l)}{2}\right) -
\ln(P(N_{\text{min}}))\right \rbrack,
\end{equation}
where $P(N_{\text{max}}^g)$ and $P(N_{\text{max}}^l)$ are the
maxima of the gas and liquid peaks, respectively, and
$P(N_{\text{min}})$ is the minimum between the two peaks. We
determine the bulk interfacial tension $\gamma_{\lg}$ by
performing simulations for a range of system sizes and by
extrapolating the results to the infinite system size using the
relation\cite{Binder1982,Potoff2000, Vink2005b}
\begin{equation} \beta \gamma_{\lg,L} =
\beta \gamma_{\lg} - \frac{x\ln L}{2 L^2}- \frac{\ln A}{2 L^2},
\label{Eq:fit1}
\end{equation}
where $A$ and $x$ are generally unknown.  The finite size scaling
is performed using the simulation results for box lengths
$L/\sigma=$8, 10, 12, and 14.

The critical temperature $T_{cr}$ and critical packing fraction
$\eta_{cr}$ are determined by fitting the scaling law
\begin{equation}
\eta_l-\eta_g=f_1 (T_{cr}-T)^{0.32}, \label{Eq:scaling}
\end{equation}
and the law of rectilinear diameters
\begin{equation}
\frac{\eta_l+\eta_g}{2}=\eta_{cr}+ f_2 (T_{cr}-T),
\label{Eq:rectilinear}
\end{equation}
to the simulation results for the gas ($\eta_{g}$) and liquid
($\eta_{l}$) packing fractions, where $f_1$ and $f_2$ are fitting
parameters.

\section{Results}
\label{S:results}

\begin{figure}[]
   \centering
   \includegraphics[height=4.0cm]{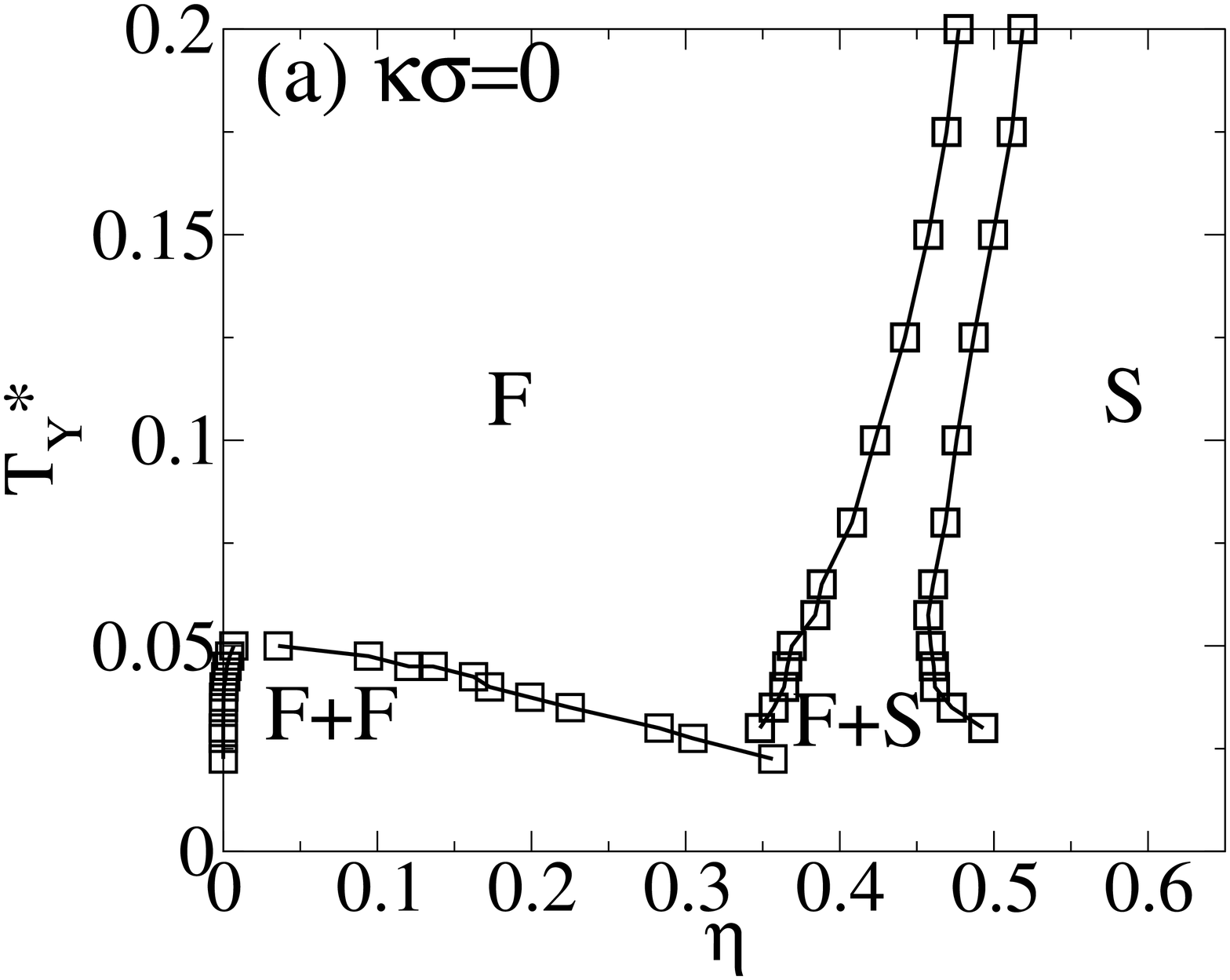}
   \includegraphics[height=4.0cm]{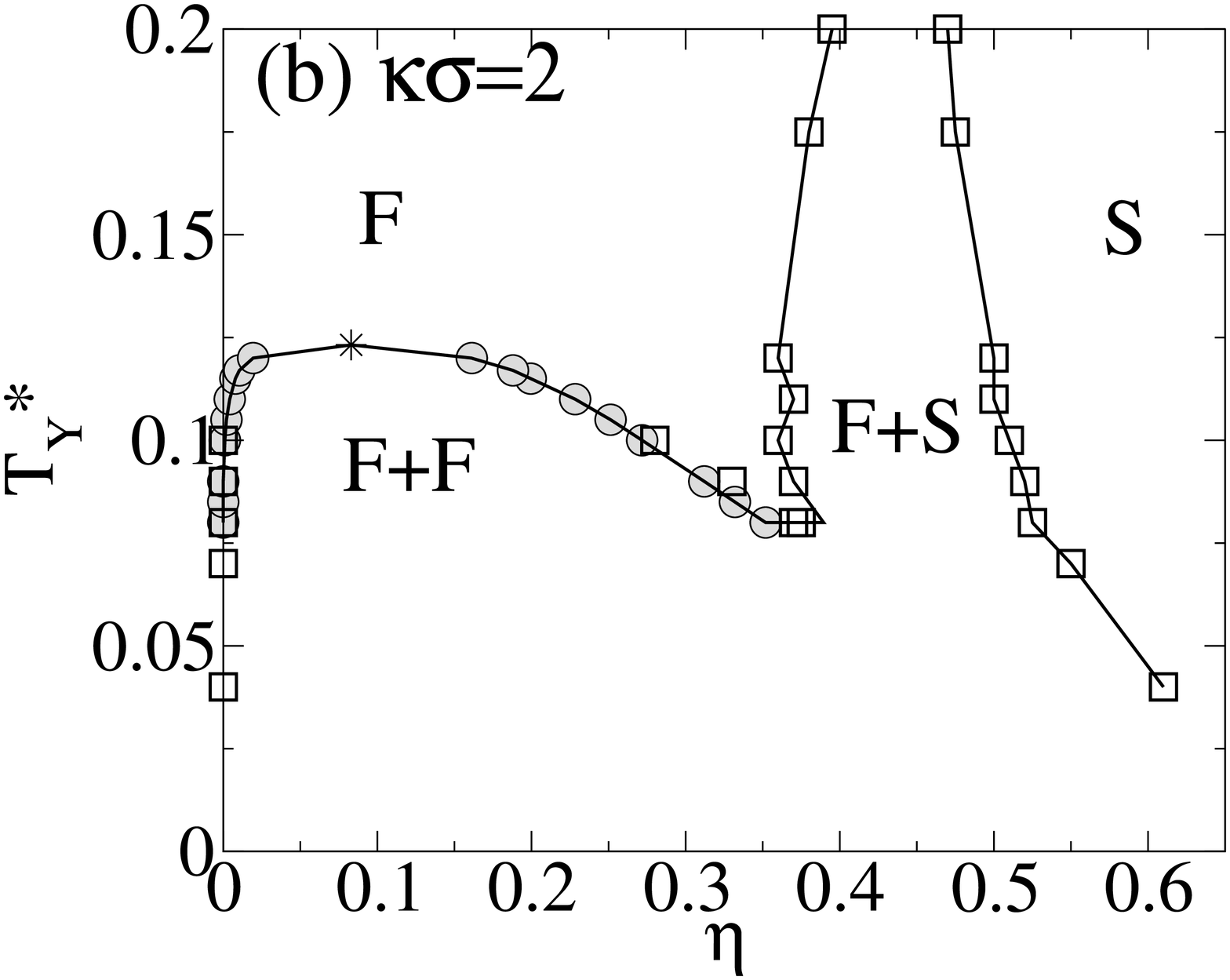}
     \includegraphics[height=4.0cm]{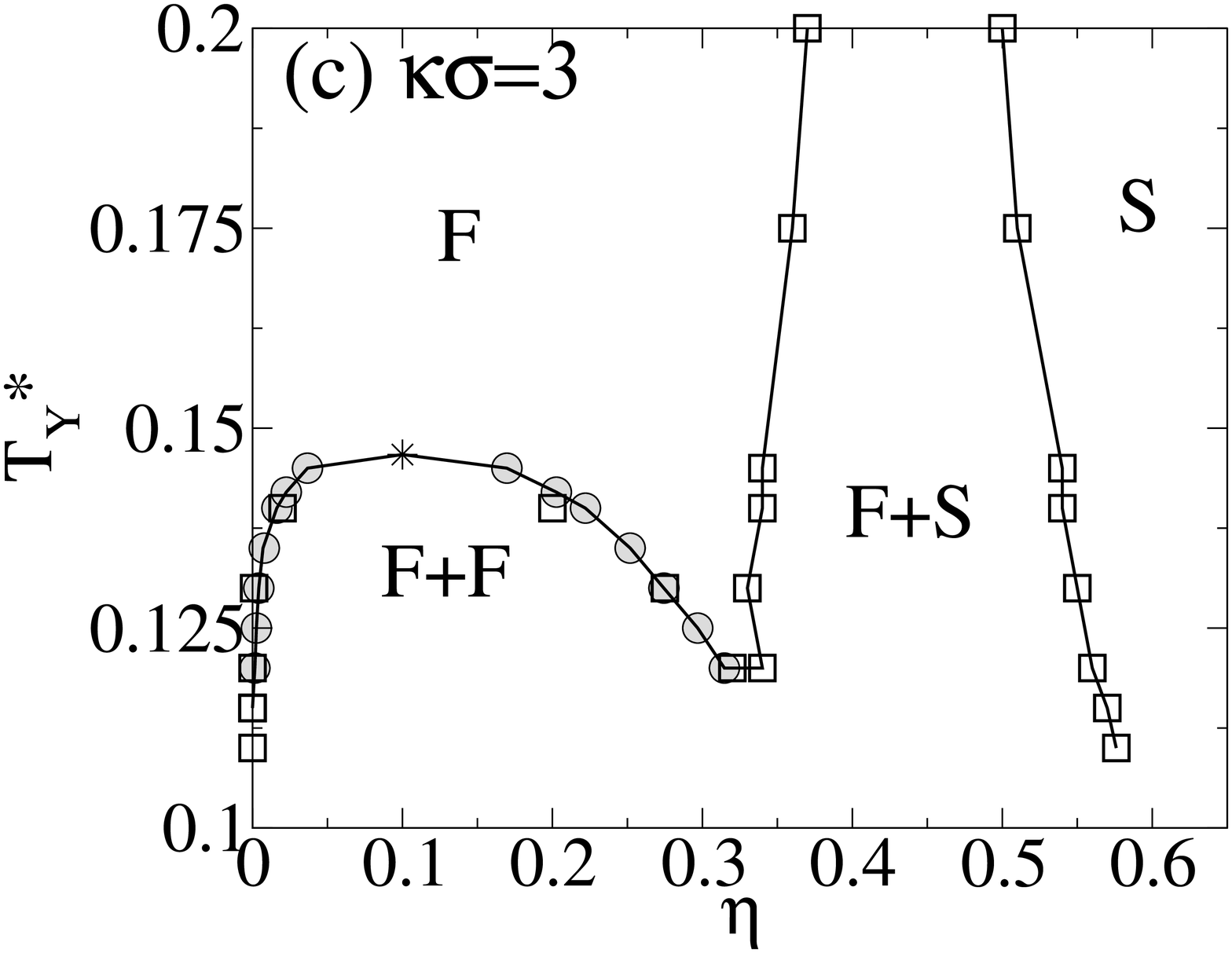}
       \includegraphics[height=4.0cm]{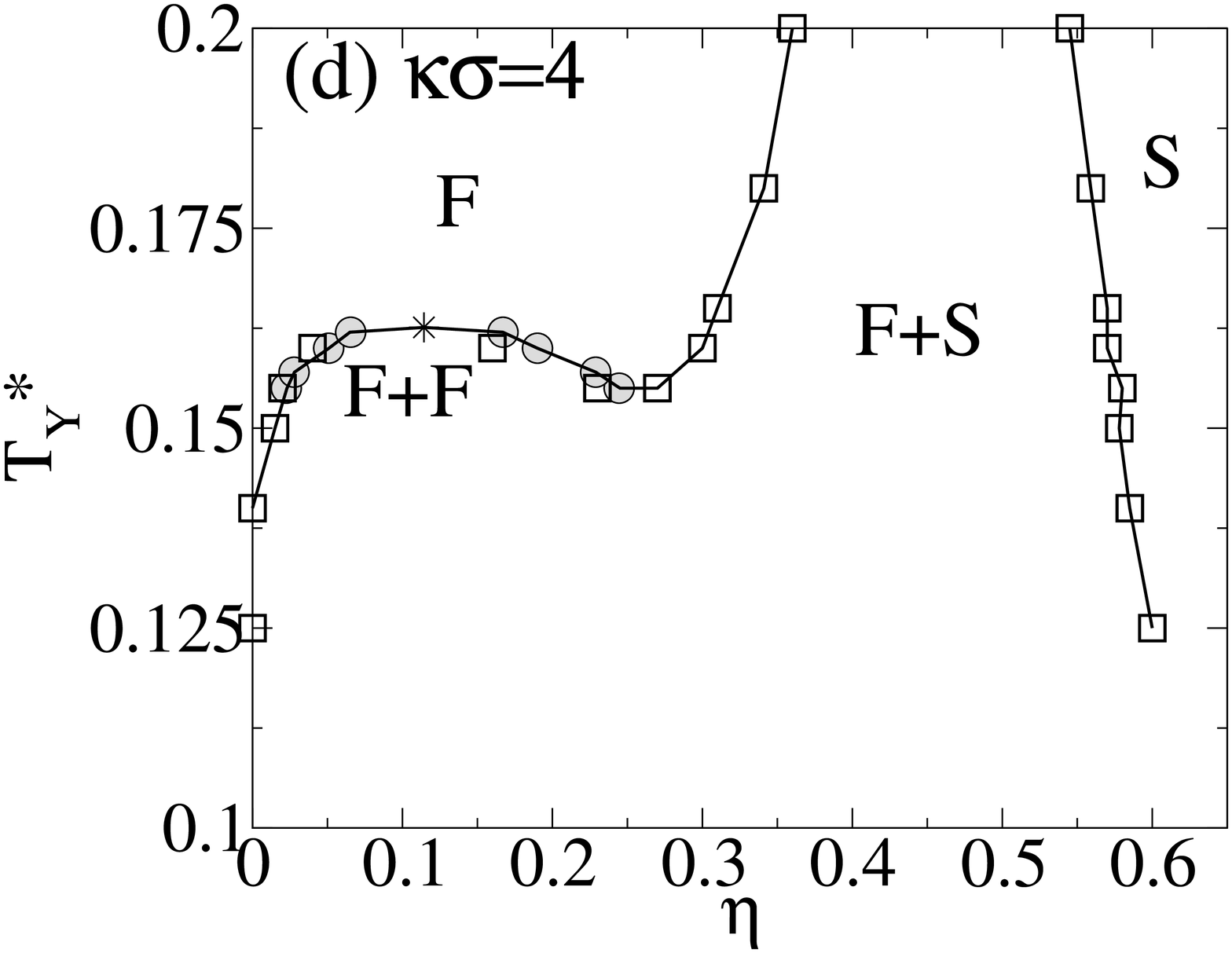}
       \includegraphics[height=4.0cm]{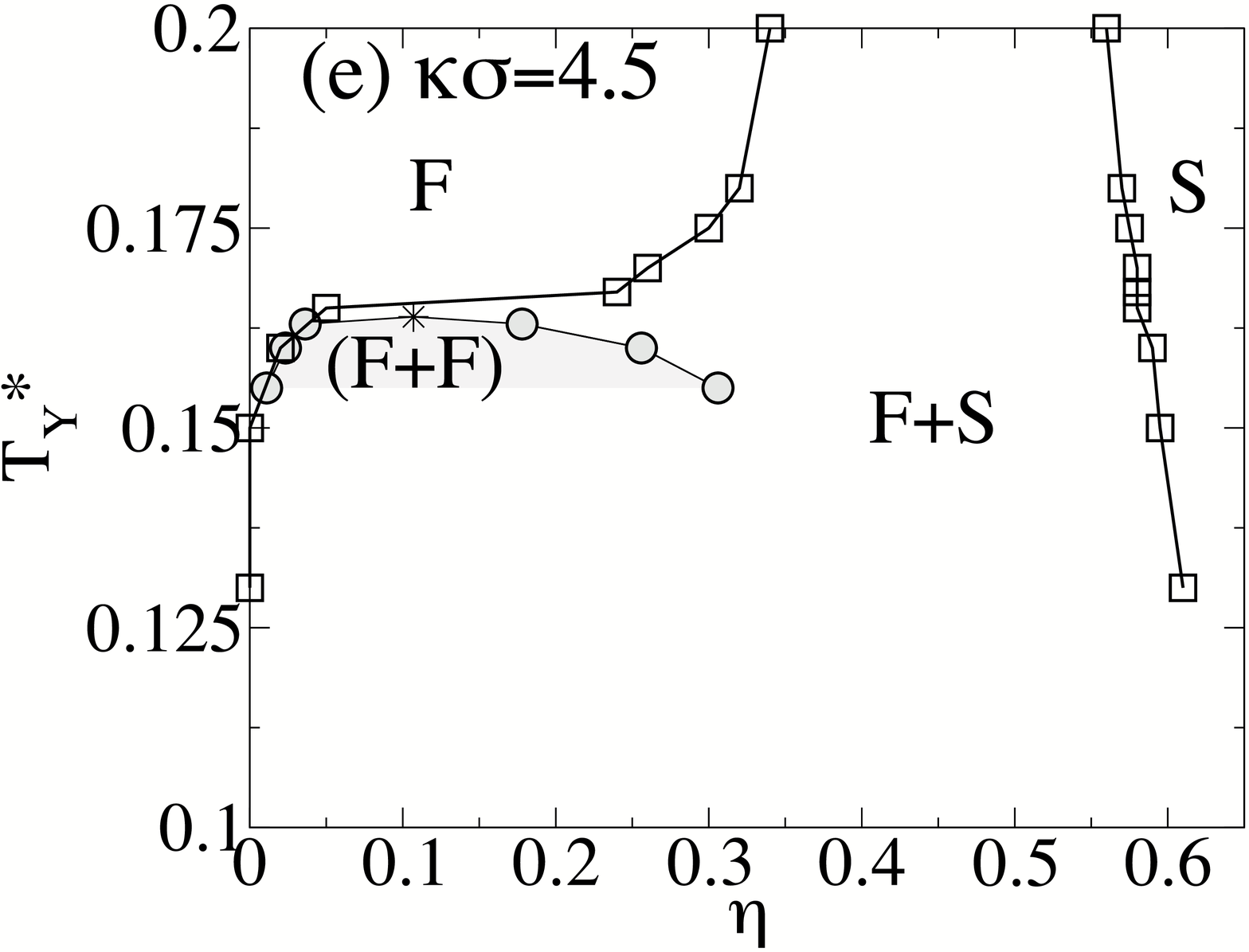}
         \includegraphics[height=4.0cm]{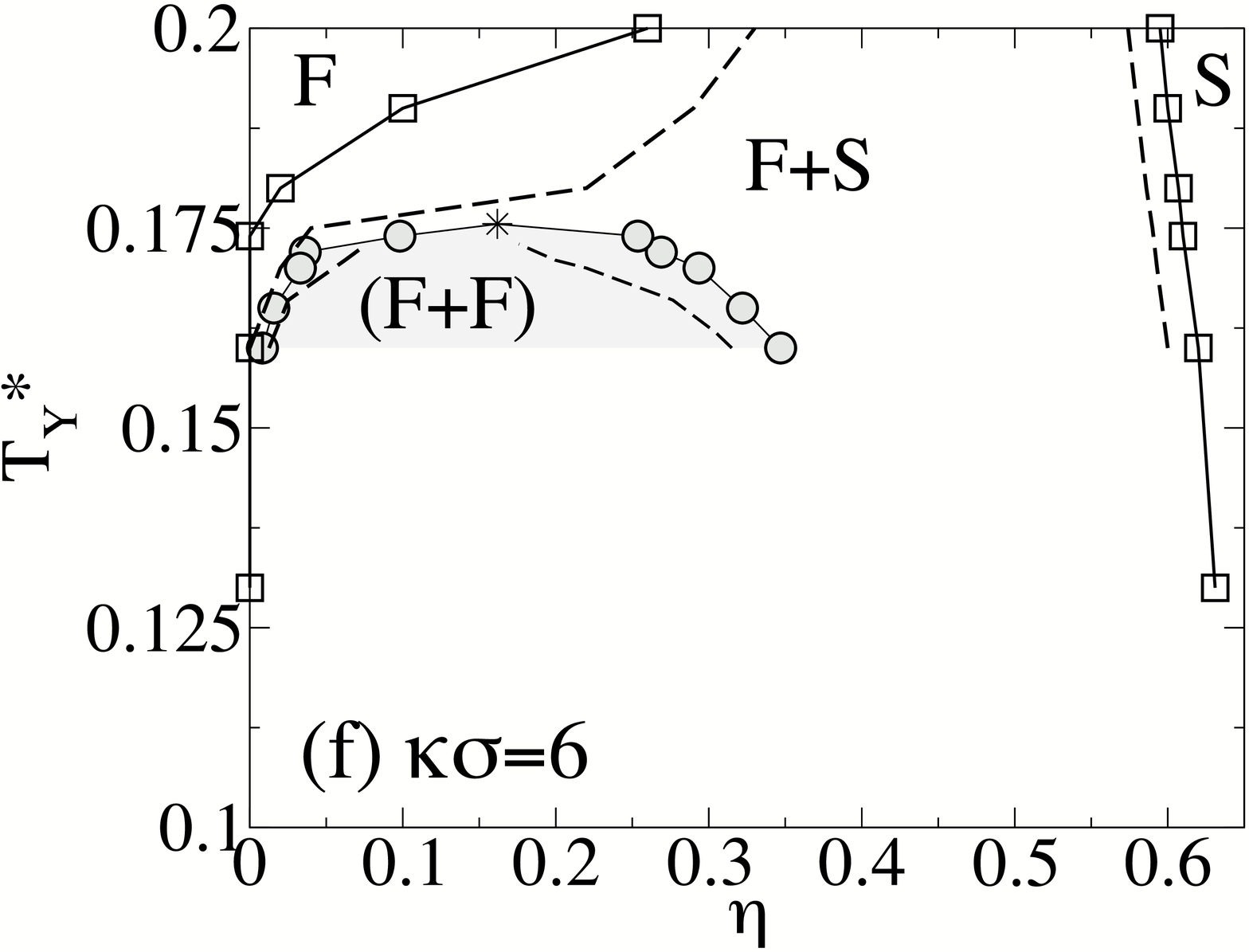}
\caption{Phase diagrams of the YRPM  in the reduced temperature
$T_{\rm Y}^*$ - packing fraction $\eta$ representation for varying
Debye screening parameter (a) $\kappa \sigma=0$ from
Refs.~\cite{Vega2003, Vega1996, Orkoulas1994}; (b) $\kappa
\sigma=2$; (c)  $\kappa \sigma$=3; (d)  $\kappa \sigma=4$; (e)
$\kappa \sigma=4.5$; (f) $\kappa\sigma=6$. The squares are the
results from the free energy calculations and the circles are the
results from the grand-canonical Monte Carlo simulations. F and S
denote the stable fluid and solid (CsCl) phase. F + S and F + F
denote, respectively, stable fluid-solid and (meta)stable
gas-liquid coexistence region (the shaded regions are metastable). The dashed line in (f) indicates
the results of the model used by~\citet{Caballero2005a}. The lines
are a guide to the eye. Tie lines (not shown) are horizontal.}
\label{fig:phds}
\end{figure}

 We compute the phase diagram using thermodynamic
integration and grand-canonical Monte Carlo simulations for
screening parameters $\kappa \sigma$=2, 3, 4, 4.5, and 6.  In Fig.
\ref{fig:phds}, we show the resulting phase diagrams in the
($\eta,T_{\rm Y}^*$) plane, together with the $\kappa \sigma$=0
phase diagram from Refs.~\cite{Vega2003} (fluid-solid) and
\cite{Vega1996, Orkoulas1994} (gas-liquid). The squares denote the
results from the free energy calculations, and the circles
represent the gas-liquid binodal obtained from the grand-canonical
Monte Carlo simulations. We find good agreement between both
results. The shaded areas in Figs. \ref{fig:phds}(e)-(f) represent
the metastable gas-liquid regions for screening parameters $\kappa
\sigma=6$ and 4.5.
For smaller screening parameters, $\kappa \sigma$=4, 3, and 2, the
gas-liquid transition is stable, and the phase diagram resembles
that of a simple fluid. At sufficiently low temperatures, a
 gas-liquid phase separation (metastable for screening parameters $\kappa
\sigma=6$ and 4.5) appears at low
densities and a fluid-solid transition at high density. At the
triple point, the gas, liquid, and the solid phase are in
coexistence, while at the critical point, the gas and the liquid
phase have the same density. At temperatures below the triple
point, a dilute gas coexists with a high density solid, and at
temperatures higher than the critical temperature, a fluid
coexists with a solid phase. Figure \ref{fig:phds} shows that the
region of stable liquid phase increases upon increasing the range
of the interaction, i.e., decreasing $\kappa \sigma$. For simple
fluids with short-range attractive Yukawa interactions,
square-well attractions, and depletion attractions, the
relationship between the range of the attractive interactions and
the stability of the gas-liquid transition has been well-studied
by computer simulations, density functional calculations, and
integral equation theories.\cite{Hagen1994, Bolhuis1994,
Tejero1994, Tejero1994b, Daanoun1994, Caccamo1999, Dijkstra2002a}
These studies show that the minimum range of attractions required
for a stable gas-liquid transition is about one sixth of the range
of the repulsions.

Our results show that the gas-liquid coexistence is stable for
$\kappa \sigma \leq 4$, and therefore contradict the findings of
Ref.~\cite{Caballero2005a}, where an estimate $\kappa \sigma \leq
25$ was given. In this comparison, we have to keep in mind that
the pair potential used in Ref.~\cite{Caballero2005a} did not
include the factor $1/r$, which we include. To study the effect of
this factor, we repeated our free energy calculations using the
pair potential of Ref.~\cite{Caballero2005a} for screening length
$\kappa \sigma=6$. The results for this model are presented in
Fig.~\ref{fig:phds}(f) with a dashed line. We find that there is
no qualitative difference between the two models; both predict a
metastable gas-liquid transition for screening length $\kappa
\sigma=6$. In Ref.~\cite{Caballero2005a}, the stability of the
gas-liquid transition was determined from the cross-over between
the freezing line and the liquid branch of the gas-liquid binodal.
The cross-over point was recognized as the gas-liquid-solid triple
point. When the triple point was at a lower temperature than the
critical point, the gas-liquid phase separation was considered
stable. Since the CsCl structure melts in the middle of the broad gas-solid coexistence, the criteria
used in Ref.\ \cite{Caballero2005a} would indicate a stable
gas-liquid separation, whereas our free energy calculations show
that it is, in fact, metastable. We argue that the computation of
the melting line cannot be used to determine the stability of
gas-liquid transition with respect to a broad gas-solid
coexistence region.

\begin{figure}[h]
\centering
\includegraphics[height=8.0cm]{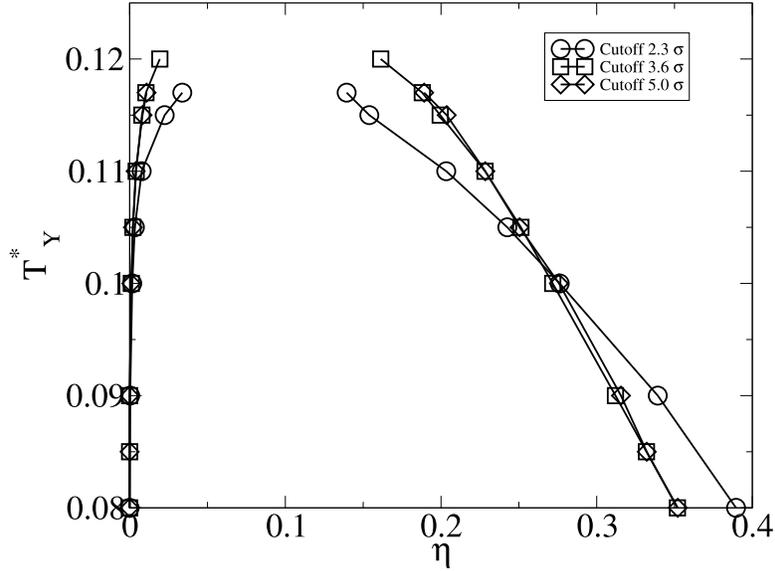}
\caption{Binodals  of the YRPM for
screening parameter $\kappa \sigma$=2, and cutoff values $r_{\text{cut}}/\sigma$=2.3, 3.6, and 5 in the
($\eta,T_{\rm Y}^*$) representation. Statistical errors are of the order of the symbol size. The lines are a guide to the eye.}
   \label{fig:cut}
\end{figure}

It is interesting to note that previous simulation studies of a one-component 
hard-core attractive Yukawa fluid predict a stable gas-liquid
transition for $\kappa \sigma=3.9$, while it is metastable for
$\kappa \sigma = 7$,\cite{Hagen1994,Dijkstra2002a} which compares
well with our results.

Table I summarizes the critical temperatures  $T_{\rm Y, cr}^*$
and critical packing fractions $\eta_{\rm cr}$ as found from our
simulations and from the generalized mean spherical approximation
(GMSA) theory\ \citep{Carvalho1997} for different values of
$\kappa \sigma$.
\begin{table}[h]
\begin{center}
\begin{tabular}{ c| c c |  c c}
\hline
 &  Simulation & & GMSA theory &  \\
 \hline
$\kappa \sigma$ & $T_{\rm Y, cr}^*$& $\eta_{\rm cr}$& $T_{\rm Y, cr}^*$ &$\eta_{\rm cr}$ \\
\hline
      6 &  0.1755 (5) & 0.162(8) & 0.16053 &0.15040 \\
      4 &  0.1626 (1) & 0.114(3) & 0.16498 & 0.11961 \\
      3 &  0.1467(1) & 0.100(1) &0.16240 & 0.09566\\
      2 &  0.1232(8) & 0.083(1) & &\\
      0 &  0.0490(3)\footnote{Data from  \citet{Orkoulas1999} } &0.037(3)&  0.07858 & 0.01448\\
\hline
\end{tabular}
\end{center}
\label{tab:crit} \caption{Critical temperatures $T_{\rm Y, cr}^*$
and packing fractions $\eta_{\rm cr}$ for the YRPM for different
values of the Debye screening parameter $\kappa \sigma$. The GMSA
theory data is from Ref.~\cite{Carvalho1997}.}
\end{table}

In Figure \ref{fig:cut}, we analyze the effect of the cutoff value of equation (\ref{Eq:yu}) on the liquid-gas binodal. Huge deviations are expected as the interaction becomes longer ranged. We used three different cutoff values $r_{\text{cut}}/\sigma=2.3, 3.6, 5$ for the calculation of the liquid-gas binodal for our longest range interaction ($\kappa \sigma=2$). The binodals for cutoff values  $r_{\text{cut}}/\sigma=3.6$, and 5 are equivalent within the statistical accuracy, thereby justifying the choice of a cutoff value $r_{\text{cut}}/\sigma=3.6$ in all subsequent calculations.

\begin{figure}[h]
\centering
\includegraphics[height=8.0cm]{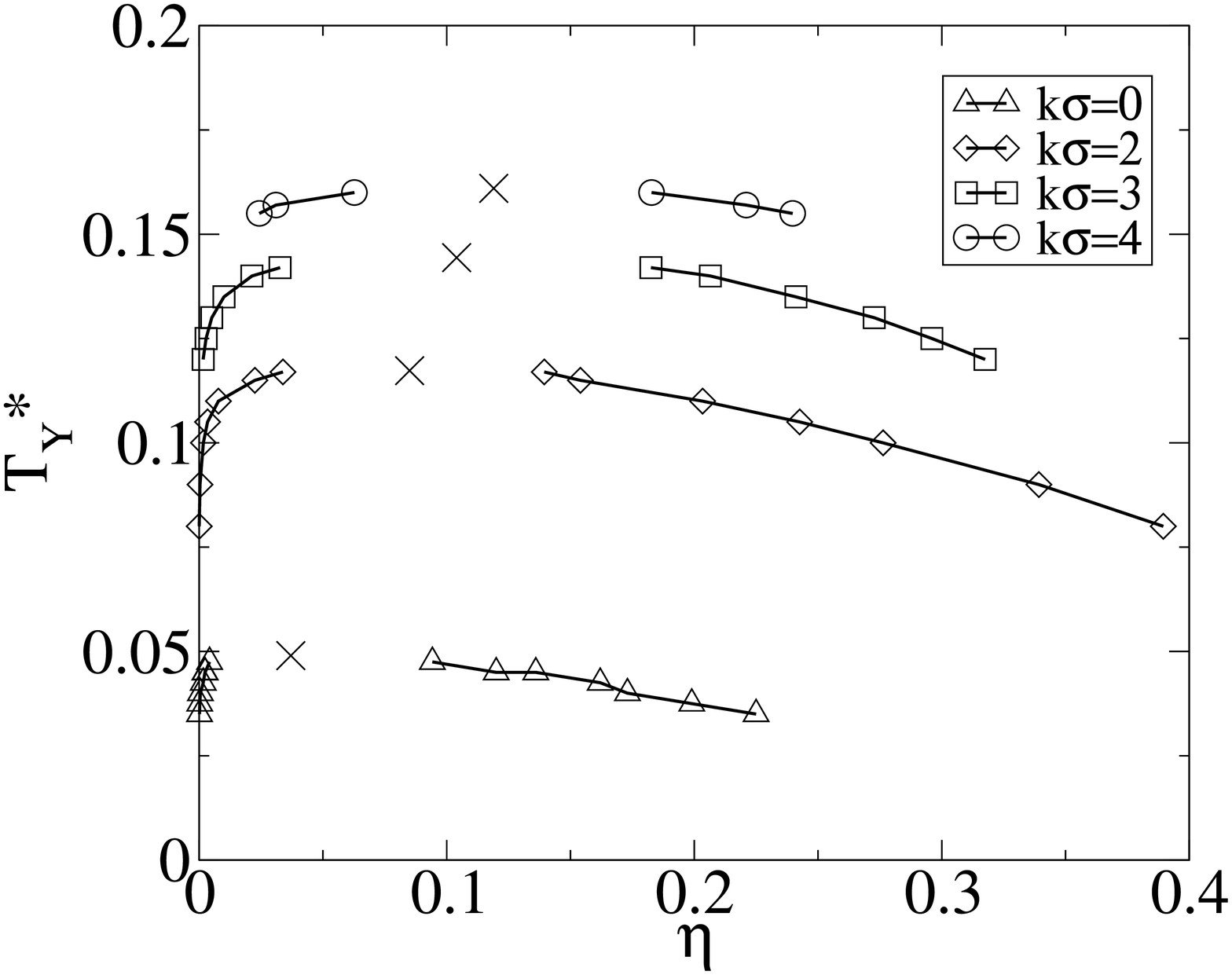}
\caption{Binodals and the critical points of the YRPM for
screening parameters $\kappa \sigma$=0, 2, 3, and 4, in the
($\eta,T_{\rm Y}^*$) representation.  Crosses denote the location of
the critical points. The binodal for $\kappa \sigma$=0 (RPM) is
taken from Ref.~\cite{Orkoulas1994} and the critical point from
Ref.~\cite{Orkoulas1999}. The lines are a guide to the eye.}
   \label{fig:binodals1}
\end{figure}

Figure \ref{fig:binodals1} shows the (stable) gas-liquid binodals for $\kappa
\sigma$=2, 3, and 4,  in the  ($\eta,T_{\rm Y}^*$) representation.
Table I and Fig. \ref{fig:binodals1} show that, for $0 \le
\kappa\sigma \le 6 $, the reduced critical temperature $T_{\rm
Y,cr}^*$ and the critical packing fraction $\eta_{\rm cr}$
decrease for increasing range of the interaction, i.e., decreasing
$\kappa \sigma$, in agreement with the findings of
\citet{Caballero2005}. The non-monotonic behavior of the critical
temperature as a function of $\kappa\sigma$ that was reported in
Ref.~\cite{Caballero2005} for screening parameters $\kappa \sigma
>10$ is in the region where we claim the gas-liquid phase
separation to be metastable. As can be seen from Table I, the GMSA
theory predicts a non-monotonic behavior of the critical
temperature as a function of $\kappa\sigma$. Comparing the
theoretical results with our simulations, we observe that the GMSA
theory overestimates the critical temperature for $\kappa \sigma <
6$ and underestimates it at $\kappa \sigma=6$. On the other hand,
the GMSA theory underestimates the critical packing fraction
$\eta_{\rm cr}$ at $\kappa\sigma=0$, but agrees reasonably well
with our simulation results for $\kappa\sigma \ge 3$.
\begin{figure}[h]
   \centering
      \includegraphics[height=6.0cm]{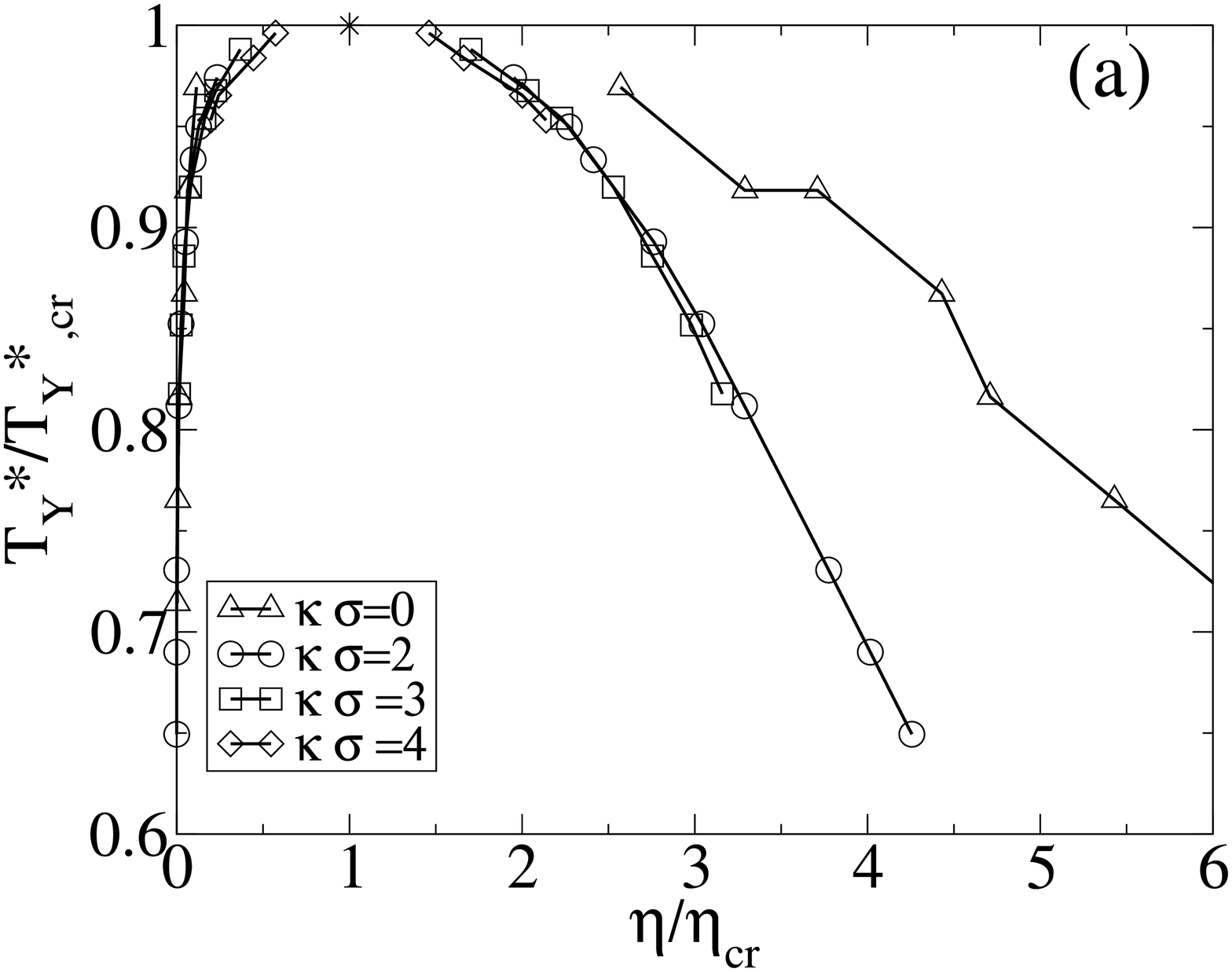}
    \includegraphics[height=6.0cm]{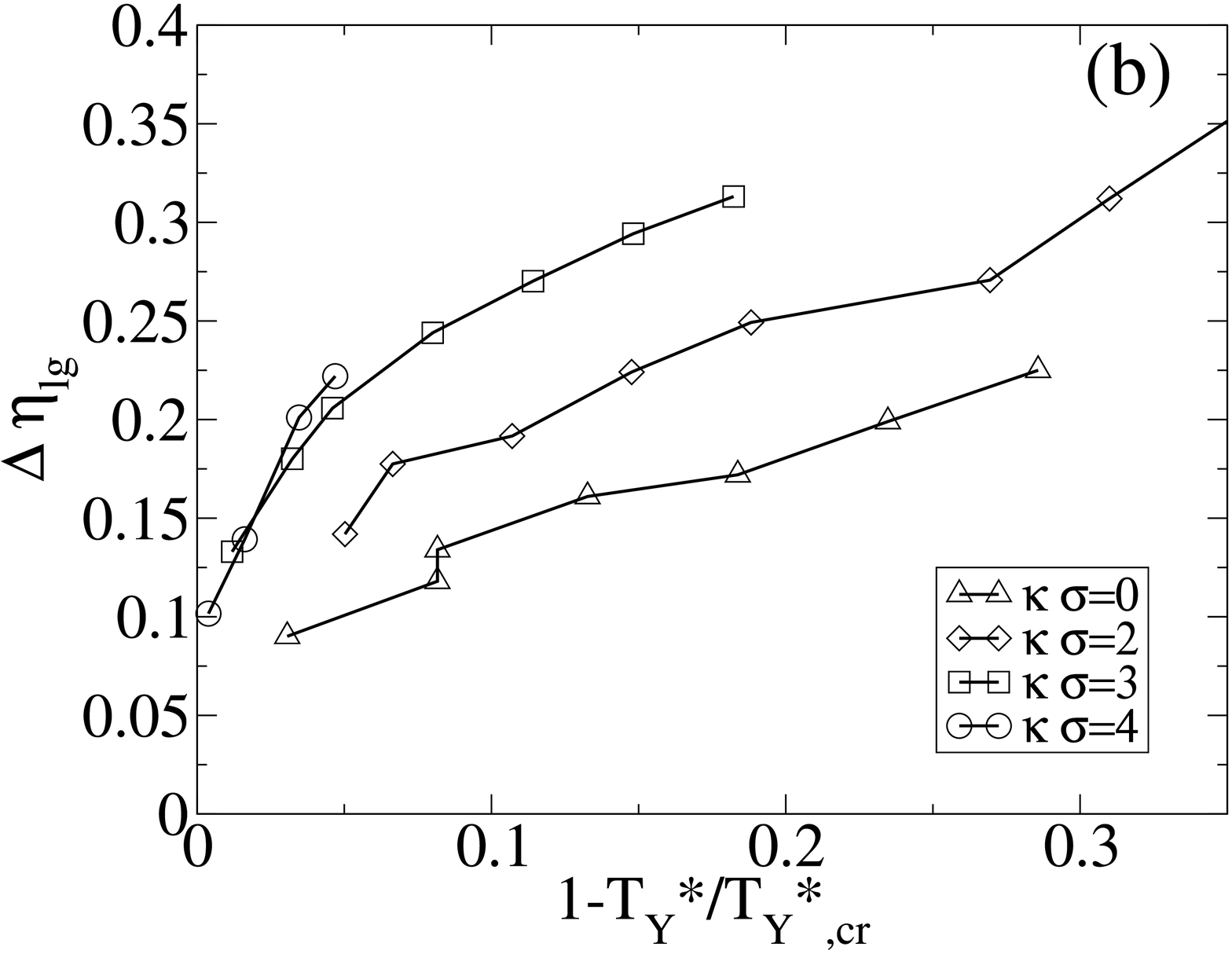}
\caption{(a) Binodals of the YRPM and the RPM in the corresponding
state representation for Debye screening parameters $\kappa
\sigma=2,3$, and 4. The reduced temperature $T_{\rm Y}^*$ is
scaled with the critical temperature $T_{\rm Y, cr}^*$ and the
packing fraction $\eta$ is scaled with the critical packing
fraction $\eta_{\rm cr}$. The binodal for $\kappa \sigma$=0 (RPM)
is from Ref.~\cite{Orkoulas1994} and the critical point from
Ref.~\cite{Orkoulas1999}. (b) The difference in coexisting packing
fractions $\Delta \eta_{\rm lg}=\eta_l-\eta_g$ is plotted against
$1-T_{\rm Y}^*/T_{\rm Y, cr}^*$.}
   \label{fig:deltarho}
\end{figure}

Figure \ref{fig:deltarho}(a) shows the binodals of the YRPM and
the RPM in the corresponding state representation, where the
reduced temperature $T_{\rm Y}^*$ is scaled with the critical
temperature $T_{\rm Y, cr}^{*}$, and the packing fraction $\eta$
is scaled with the critical packing fraction $\eta_{\rm cr}$. We
see that the binodals do not collapse on a single master-curve,
but instead, the RPM binodal (where $\kappa\sigma=0$) differs
considerably from the YRPM binodals (where $\kappa \sigma=2,3$,
and 4). This finding is in agreement with the prediction of the
GMSA theory.\cite{Carvalho1997} 
In Fig.\ \ref{fig:deltarho}(b), we plot the width of the gas-liquid separation, $\Delta \eta_{\rm lg}=\eta_l-\eta_g$, as a function of $1-T_{\rm Y}^*/T_{\rm Y,
cr}^*$. We see that for a fixed $1-T_{\rm Y}^*/T_{\rm Y, cr}^*$,
the width of the gas-liquid separation decreases with increasing
range of the interaction, resulting in a smaller density gap
between the coexisting liquid and gas phase for longer-ranged
interactions.

\begin{figure}[h]
   \centering
   \includegraphics[height=8cm]{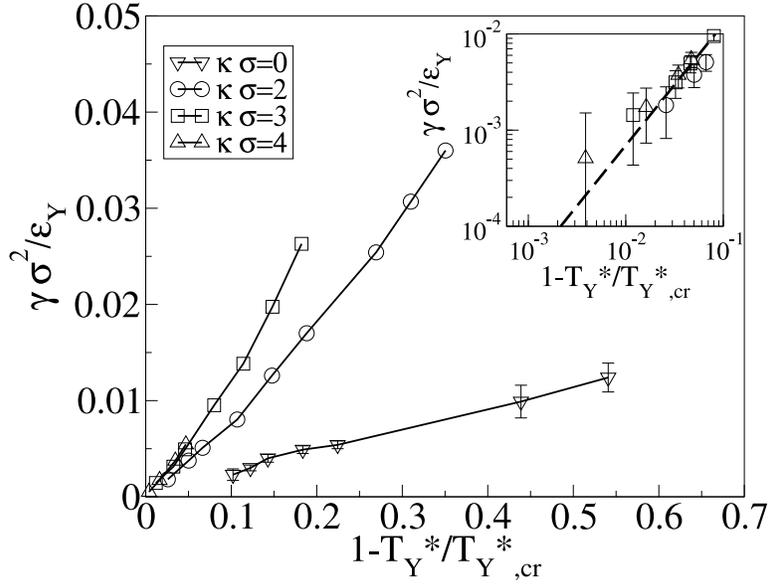}
\caption{Dimensionless gas-liquid interfacial tension $\gamma_{\rm
lg}\sigma^2/\epsilon_{\rm Y}$ as a function of $1-T_{\rm Y,
cr}^*/T_{\rm Y}^*$. The data for $\kappa\sigma=0$ is from
Ref.~\cite{Gonzalez-Melchor2003}. 
Inset: log-log plot of the dimensionless interfacial tension in the vicinity of the critical point.
The dashed line is the theoretical prediction\cite{Chen1982,Ferrenberg1991} $\gamma \sim (T_{\rm Y, cr}^*-T_{\rm Y}^*)^{1.26}$.}
 \label{fig:gammadimless}
\end{figure}

Figure \ref{fig:gammadimless} shows the gas-liquid interfacial
tension, scaled with the contact value energy $\epsilon_{\rm Y}$,
for different values of the screening parameter $\kappa \sigma$.
For comparison, we also show the interfacial tension of the RPM
from Ref.~\cite{Gonzalez-Melchor2003}. As can be seen from Fig.\
\ref{fig:gammadimless}, the value of the dimensionless interfacial
tension increases with increasing $\kappa\sigma$. This can be
understood on the basis of Fig.\ \ref{fig:deltarho}(b), which
shows that, with increasing $\kappa\sigma$, the density gap of the
coexistence region increases, meaning that the interfacial tension
increases.
The inset of figure \ref{fig:gammadimless} shows a log-log plot of  $\gamma_{\rm
lg}\sigma^2/\epsilon_{\rm Y}$ versus  $1-T_{\rm Y}^*/T_{\rm Y,
cr}^*$ in the vicinity of the critical point, which can be used to extract an estimate for the critical exponent of the correlation length $\nu$. 
We found $2 \nu \simeq 1.1$ for all screening parameters $\kappa\sigma$, by performing a linear fit on the data, 
which differs from the Ising model result $2 \nu = 1.32$,\cite{Binder1982} and from the accepted value of $2 \nu =1.26$.\cite{Chen1982,Ferrenberg1991}
 The value of the correlation length is very sensitive to the extrapolated surface tension to infinite system sizes. 
In order to improve the statistical accuracy of the simulations, larger system sizes, as well as 
longer runs are needed, especially close to the critical point. Nevertheless, our results are compatible within the simulation error, with the theoretical prediction of the correlation length (dashed line in the inset of figure  \ref{fig:gammadimless}).

\begin{figure}[h]
\centering
\includegraphics[height=6cm]{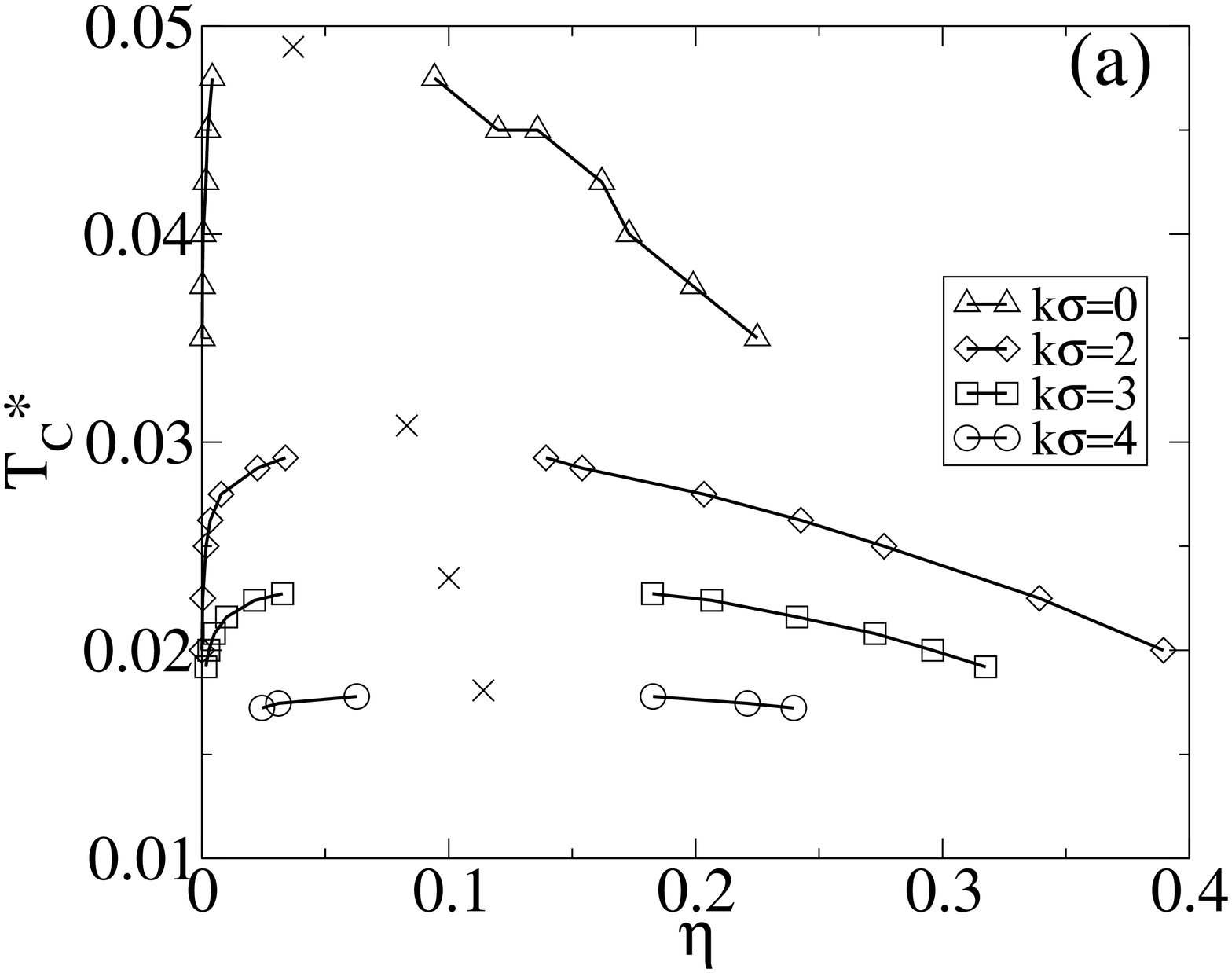}
\includegraphics[height=6cm]{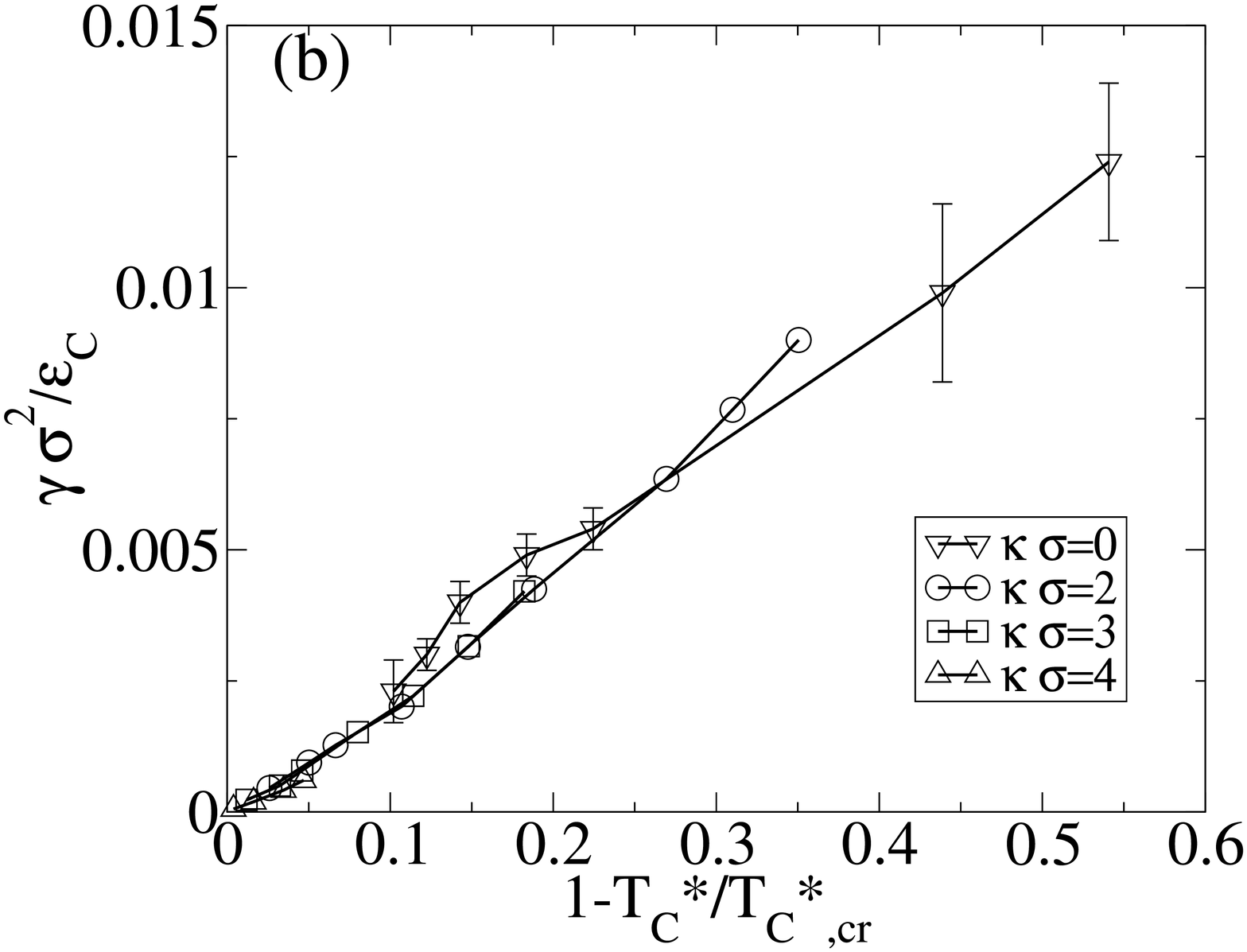}
\caption{(a) Binodals of the YRPM, for Debye screening parameters
$\kappa \sigma=$0, 2, 3, and 4, in the reduced temperature $T_{\rm
C}^*$  and the packing fraction $\eta$ representation. The binodal
for $\kappa \sigma$=0 (RPM) is from Ref.~\cite{Orkoulas1994} and
the critical point from Ref.~\cite{Orkoulas1999}. (b)
Dimensionless gas-liquid interfacial tension $\gamma_{\rm
lg}\sigma^2/\epsilon_{\rm C}$ as a function of $1-T_{\rm
Y}^*/T_{\rm Y, cr}^*$. Remember that $T_{\rm C}^*/T_{\rm C,
cr}^*=T_{\rm Y}^*/T_{\rm Y, cr}^*$. The data for $\kappa\sigma=0$
is from Ref.~\cite{Gonzalez-Melchor2003}.}
   \label{fig:phdrpm}
\end{figure}

We now interpret our results in view of the DLVO theory. In the
DLVO theory, the contact value $\epsilon_Y$, and hence the reduced
temperature $T_{\rm Y}^*=k_B T/\epsilon_{\rm Y}$, depend on the
salt concentration $\rho_s$ through the screening parameter
$\kappa \sigma$, see Eq.~(\ref{dlvo}). In Fig.\
\ref{fig:phdrpm}(a), we plot the gas-liquid binodals and critical
points for $\kappa \sigma$=0, 2, 3, and 4, using the reduced
temperature $T_{\rm C}^*=\sigma/Z^2 \lambda_B$ that does not
depend on $\kappa \sigma$.
As can be seen from Fig. \ref{fig:phdrpm}(a), the reduced critical
temperature  $T_{C, cr}^*$  decreases with increasing $\kappa
\sigma$, or salt concentration. This means that, at a fixed
$T^*_C$ and at a statepoint inside the gas-liquid coexistence
region, adding salt decreases the density difference between the
gas and the liquid phases, until, at the critical salt
concentration, the density difference disappears. This finding
could be confirmed by performing simulations with explicit co- and
counterions, \cite{Hynninen2005, Hynninen2005b} and could be used
to experimentally test the validity of the DLVO theory for
oppositely charged colloids. Figure \ref{fig:phdrpm}(b) shows the
interfacial tension scaled with the contact value 
$\epsilon_{\rm C}=\frac{ Z^2 k_B T \lambda_B}{ \sigma}$, and as can be seen, the interfacial tensions for different
$\kappa\sigma$ collapse to a single line. This suggests that the
interfacial tension is determined solely by the contact value and
not by the range of the interaction.

\section{Conclusions}
We have used a combination of MC free energy calculations and
grand-canonical MC simulations to determine the stability and the interfacial
tension of the gas-liquid phase separation in a binary mixture of
oppositely charged hard spheres, which interact via
screened-Coulomb (Yukawa) potentials. We find that the gas-liquid
coexistence is stable with respect to gas-solid coexistence for
values of the screening parameter $\kappa \sigma \leq 4$. This
value is similar to what is found for the single component
attractive Yukawa model,\cite{Dijkstra2002a} where the gas-liquid
transition is stable at $\kappa \sigma=4$ and metastable at $\kappa \sigma=7$.

We have studied the dependence of the critical temperature as a
function of the range of the Yukawa interaction. If the contact
value of the interaction potential does not depend on the
screening length, it is possible to define a reduced critical
temperature simply as the inverse of the Yukawa contact value.
With this definition, the reduced critical temperature decreases
upon increasing the range of the interaction, which is in
agreement with Ref.\ \cite{Caballero2005}.

We have related the Yukawa restricted primitive model (YRPM) to
the DLVO theory, which was recently used to explain experimental
results on oppositely charged colloids.\cite{Leunissen2005,
Hynninen2006, Hynninen2006b, Maskaly2006} The DLVO theory
predicts a contact value that depends on the screening length.
Thus, in order to facilitate the comparison between the results
for different screening lengths, we define a temperature scale
that is independent of the screening length. The natural choice is
the reduced temperature of the RPM, which is the limit of zero
screening length of the DLVO theory. With this definition, the
reduced critical temperature decreases upon increasing the range
of the interaction. This means that upon adding salt to a system
at fixed temperature and at a statepoint in the gas-liquid
coexistence region, the density difference between the gas and
liquid phases decreases, and finally disappears at the critical
salt concentration. This prediction could be tested by computer
simulations with explicit co- and counterions,\cite{Hynninen2005,
Hynninen2005b} and could be used to study experimentally the
validity of the DLVO theory for oppositely charged colloids.

Finally, we have studied the gas-liquid interfacial tension using
histogram reweighting methods. We find that the dimensionless tension decreases 
for decreasing screening parameter. Upon scaling the
interfacial tension with the contact value of the Coulomb
interaction, we observed a collapse of the interfacial tensions
onto a single curve. This means that for state points at
coexistence and at the same scaled temperature $T_{\rm C}^*/T_{\rm
C, cr}^*$, the interfacial tension is determined solely by the
contact value and not by the range of the interaction. There
might be a possible connection with the well known similarities between
the structures of the RPM and of the YRPM. \citet{Larsen1980}
noted that their Monte Carlo results for the radial distribution
functions of the YRPM were similar to those obtained for the RPM
at different state points. The structurally equivalent states were
further investigated by \citet{Copestake1982} and by Leote de
Carvalho and Evans \cite{Carvalho1997}. This correspondence is due
to the screening of charges in the RPM, which in turn is due to
charge ordering.\cite{Larsen1980,Copestake1982,Carvalho1997}
Consequently, the potential of mean force between two ions decays
more rapidly than the bare Coulomb potential. For certain state
points the potentials of mean force will be similar for the YRPM
and the RPM. Preliminary results suggest that a similar
explanation holds for the state points with similar interfacial
tensions. Further investigation is in progress.

\acknowledgments We thank  A. Cuetos and E. Sanz for many useful
and inspiring discussions. We thank Athanassios Z.
Panagiotopoulos for critical reading of the manuscript. 
This work is part of the research program of the {\em Stichting voor Fundamenteel Onderzoek der
Materie} (FOM), that is financially supported by the {\em
Nederlandse Organisatie voor Wetenschappelijk Onderzoek} (NWO). We
thank the Dutch National Computer Facilities foundation for
granting access to the LISA supercomputer.

\bibliographystyle{apsrev}
\bibliography{ref}
\end{document}